\documentclass[12pt]{iopart}
\usepackage{iopams}
\usepackage{amsfonts}
\usepackage{latexsym}
\usepackage{graphicx}

\newcommand{\mean}[1]{\langle{#1}\rangle}

\newcommand{\bra}[1]{\langle{#1}|}
\newcommand{\ket}[1]{|{#1}\rangle}

\begin{document}

%%%%%%%%%%%%%%%%%%%%%%%%%%%%%%%%%%%%%%%%%%%%%%%%%%%%%%%%%%%%%%%%%%%
%%%%%%%%%%%%%%%%%%%%%%%%%%%%% Title %%%%%%%%%%%%%%%%%%%%%%%%%%%%%%%
%%%%%%%%%%%%%%%%%%%%%%%%%%%%%%%%%%%%%%%%%%%%%%%%%%%%%%%%%%%%%%%%%%%

\title[]
{Structure identification and state initialization of 
spin network with limited access}

\author{Yuzuru Kato and Naoki Yamamoto}

\address{
Department of Applied Physics and Physico-Informatics, 
Keio University, Hiyoshi 3-14-1, Kohoku-ku, Yokohama 223-8522, Japan }
\ead{yuzuru.kato.7@gmail.com, yamamoto@appi.keio.ac.jp}

%%%%%%%%%%%%%%%%%%%%%%%%%%%%%%%%%%%%%%%%%%%%%%%%%%%%%%%%%%%%%%%%%%%%%%%%%
%%%%%%%%%%%%%%%%%%%%%%%%%%%%%%%%  Abstract  %%%%%%%%%%%%%%%%%%%%%%%%%%%%%
%%%%%%%%%%%%%%%%%%%%%%%%%%%%%%%%%%%%%%%%%%%%%%%%%%%%%%%%%%%%%%%%%%%%%%%%%

\begin{abstract}

For reliable and consistent quantum information processing 
carried out on a quantum network, the network structure must 
be fully known and a desired initial state must be accurately 
prepared on it. 
In this paper, for a class of spin networks with its single 
node only accessible, we provide two continuous-measurement-based 
methods to achieve the above requirements; 
the first one identifies the unknown network structure with high 
probability, based on continuous-time Bayesian update of the graph 
structure; 
the second one is, with the use of adaptive measurement technique, 
able to deterministically drive any mixed state to a spin coherent 
state for network initialization.

\end{abstract}

%\pacs{03.67.-a, 03.65.Yz, 02.30.Yy, 42.50.Lc}
%\submitto{\NJP}

%\maketitle

%%%%%%%%%%%%%%%%%%%%%%%%%%%%%%%%%%%%%%%%%%%%%%%%%%%%%%%%%%%%%%%%%%%%%
%%%%%%%%%%%%%%%%%%%%%%%%% I. Introduction %%%%%%%%%%%%%%%%%%%%%%%%%%%
%%%%%%%%%%%%%%%%%%%%%%%%%%%%%%%%%%%%%%%%%%%%%%%%%%%%%%%%%%%%%%%%%%%%%

\section{Introduction}

Quantum information processing is usually performed on a highly 
networked system composed of many subsystems 
\cite{Divincenzo 2000, Nielsen Book}. 
In particular, the universal quantum computation is possible if 
we can ideally control {\it all} those subsystems 
\cite{Deutsch 1989, DiVincenzo 1995, Lloyd 1995, Barenco 1996}. 
However, this ``global control" approach inevitably introduces 
noise to the controlled subsystems, which accumulate and as a 
result can largely degrade the performance of information processing. 
Also engineering all the control actuators costs a lot for a large 
network. 
A different view is that all the network components are usually not 
accessible, such as a solid network system whose surface can only 
be manipulated or measured; 
hence in this case the global control approach cannot be taken. 
These facts thus stimulate development of methodologies dealing 
with networks that allow access only to a small set of subsystems.

Now let us turn our attention to the requirements imposed on 
a network for quantum computation \cite{Divincenzo 2000}. 
In particular, the followings are critical; 
the dynamics of the whole network must be fully known, a desired 
initial state of the network must be accurately prepared, and 
universal gate operation is possible. 
Together with the fact mentioned in the first paragraph, we are 
thus reasonably motivated to develop a scheme for achieving these 
requirements in a network that allows us access to only a part of 
the whole system. 
Actually there have been notable progress along this direction; 
Refs.~\cite{Burgarth 2009 1, Burgarth 2009 2,Franco 2009,Burgarth 2011, 
Lapasar 2012, Maruyama 2012} deal with the problem estimating the 
parameters of a limited-access spin network with {\it known} structure 
(topology); 
also we find a {\it probabilistic} system initialization method in 
\cite{Nakazato 2003,Nakazato 2004}; 
moreover, there have been developed some approaches to quantum 
computation, in terms of controllability analysis, in 
\cite{Schirmer 2008, Schirmer 2008 2, Burgarth 2009 3, Burgarth 2010, 
Wang 2012}.

\begin{figure} [!t]
\begin{center}
\includegraphics[width=12cm,clip]{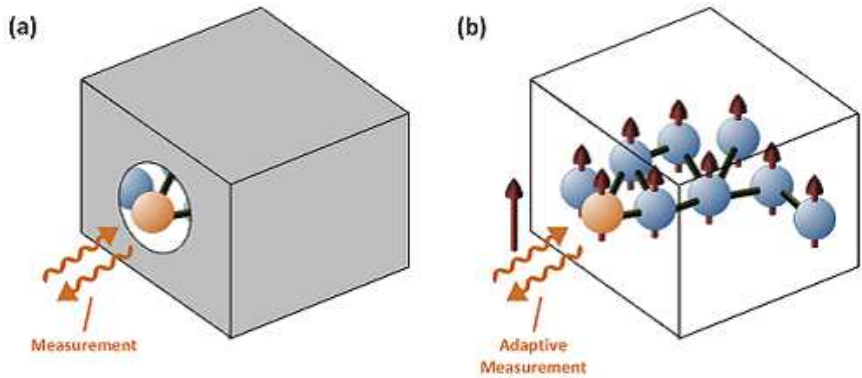}
\caption{
(a) Structure identification of an unknown spin network hidden in 
a black box. 
(b) State initialization of a known spin network. 
The orange colored node indicates the accessible spin, which is 
continuously measured. 
}
\label{fig_1}
\end{center}
\end{figure}

The purpose of this paper is to complete the procedures necessary 
for quantum computation on a limited-access network. 
That is, we aim to develop methods achieving (i) structure 
identification and (ii) deterministic state initialization, for spin 
networks with only a single node accessible. 
Indeed accomplishment of these tasks together with the results 
referred above, i.e. parameter estimation \cite{Burgarth 2009 1, 
Burgarth 2009 2,Franco 2009,Burgarth 2011, Lapasar 2012, Maruyama 2012} 
and universal gate operation \cite{Burgarth 2009 3, Burgarth 2010}, 
enable us to carry out quantum computation on a limited-access network. 
A cartoon illustrating our tasks is given in Fig.~\ref{fig_1}, where 
in the case (i) we try to identify the structure of an unknown 
network hidden in a ``black box", and in the case (ii) we try to 
stabilize a spin coherent state of a known network. 
Both of the schematics are based on {\it continuous measurement} 
\cite{Belavkin, Bouten, Wiseman Book}, which is a continuous-time 
repetition of Bayesian update of the system state based on the 
measurement result. 
Thus, it can be directly applied to the problems of parameter 
estimation \cite{Wiseman Book, Mabuchi 1996, Gambetta 2001, 
Chase 2009, Ralph 2011, Molmer 2013} and feedback control for 
state preparation \cite{Wiseman Book,Stockton 2004,Handel 2005,
Yamamoto 2007,Mirrahimi}, implying suitability of the continuous 
measurement approach in our case as well.

In what follows we describe the significance of the goals (i) and (ii) 
in more detailed way and also the results briefly. 
First, regarding the goal (i), we are motivated by the fact that 
the graph structure itself is often unknown; for instance, in the 
case of solid systems, subsystems are served by atoms produced at 
different sites possibly randomly and only some of them appear near 
the surface. 
Hence in general it is important to develop a scheme for identifying 
the graph structure from only accessible nodes, as illustrated in 
Fig.~\ref{fig_1}~(a). 
In this paper, towards achieving the goal~(i), we provide an 
algorithm to test whether any given pair of nodes of the network 
are connected or not; 
then it will be demonstrated numerically that our scheme correctly 
identifies the graph structure with high success probability. 
It is worth noting that this kind of structure identification 
problem can be found in the classical regime, e.g., reconstruction 
of the graph structure of gene mRNA concentrations \cite{Rice 2005} 
and estimation of relationships in social networks 
\cite{Siciliano 2012}.

Once the network structure is correctly identified, which means that 
the system parameters can be further estimated using the results in 
\cite{Burgarth 2009 1, Burgarth 2009 2,Franco 2009,Burgarth 2011, 
Lapasar 2012, Maruyama 2012}, then the next step is to initialize 
the network. 
In this paper, we aim to {\it deterministically} stabilize a 
spin coherent state $\ket{0^{\otimes N}}=\ket{0, 0, \cdots, 0}$ 
in a limited-access network originally prepared in {\it any} 
mixed state. 
There has not been developed a scheme satisfying all these 
requirements, though it is clearly an important subject particularly 
in quantum computation. 
The continuous measurement is again useful to solve the problem, 
because it continuously reduces the entropy of the system; 
further, to overcome the issue that any measurement induces 
probabilistic (i.e. non-deterministic) behavior of the system, we 
employ the {\it adaptive measurement} technique \cite{Jacobs 2010,
Tanaka 2012}, which is a kind of feedback control that changes the 
measured observable continuously in time, depending on the past 
measurement results. 
It will be actually demonstrated that, in some examples, this method 
realizes deterministic stabilization of the target state. 
At the same time, we clarify a situation where the adaptive scheme 
does not work; 
more precisely, it is proven that the {\it permutation symmetry} 
property \cite{Wang 2012} of the network prohibits such desirable 
convergence.

Lastly we remark that the presented schemes can be straightforwardly 
extended to the case where {\it two} neighboring nodes of the 
network are accessible; 
this is indeed a necessary requirement to perform universal quantum 
computation on a limited-access network \cite{Burgarth 2010}.

Notation: 
The spin-up and spin-down states are represented by 
$\ket{0}=(1, 0)^\top$ and $\ket{1}=(0, 1)^\top$, respectively. 
$I_n$ is the $n\times n$ identity matrix. 
$\sigma^x, \sigma^y$, and $\sigma^z$ are the Pauli matrices.

%%%%%%%%%%%%%%%%%%%%%%%%%%%%%%%%%%%%%%%%%%%%%%%%%%%%%%%%%%%%%%%%%%%
%%%%%%%%%%%%%%%%%%%%%%%%%%%%  SEC.II  %%%%%%%%%%%%%%%%%%%%%%%%%%%%%
%%%%%%%%%%%%%%%%%%%%%%%%%%%%%%%%%%%%%%%%%%%%%%%%%%%%%%%%%%%%%%%%%%%

\section{Spin network under continuous measurement}

We first describe the general setup of continuous measurement. 
This can be physically realized by coupling an optical probe 
field to the system of interest and measuring the output field 
continuously in time. 
In particular when a homodyne detector is used for measurement, 
the time evolution of the system state $\rho_t$ conditioned on 
the measurement results ${\cal Y}_t=\{Y_s~|~0\leq s \leq t\}$ 
is given by the following {\it stochastic master equation} (SME) 
\cite{Belavkin, Bouten, Wiseman Book}: 
\begin{eqnarray}
\label{eq_sme1}
& & \hspace*{0em}
     d \rho_{t} = -i[H,\rho_{t}]dt 
       + \gamma  \mathcal{D}[c]\rho_{t}dt 
        + \sqrt{\gamma} \mathcal{H}[c]\rho_{t}dW_{t}, 
 \\ & & \hspace*{0em}
\label{eq_sme2}
     dY_{t} 
      = \sqrt{\gamma}\hspace{0.05cm}{\rm Tr}[(c+c^{\dag})\rho_{t}]dt 
         + dW_t,
\end{eqnarray}
where $H$ is the system Hamiltonian. 
The {\it measurement operator} $c$ represents the coupling 
between the system and the probe field, and $\gamma$ is the 
measurement strength. 
$dW_{t}$ is the standard Winner increment with mean zero and 
variance $dt$. 
Also we have defined 
\[
    \mathcal{D}[c]\rho
        =c \rho c^{\dag}-\frac{1}{2}(c^{\dag}c\rho + \rho c^{\dag} c),~~~
    \mathcal{H}[c]\rho = c \rho + \rho c^{\dag}
         - \Tr[(c + c^\dagger) \rho]\rho. 
\]
Note that the set of equations \eref{eq_sme1} and \eref{eq_sme2} is 
a quantum counterpart to the classical Kushner-Stratonovich equation 
describing the time evolution of a conditional probability density. 
Hence, as in the classical case, the conditional expectation 
$\Tr(A\rho_t)$ represents the least mean squared error estimate 
of an observable $A$ at time $t$.

In this paper, we study an $N$-spins network whose structure is 
captured by the graph $G$ with the set of nodes (vertices) $V(G)$ 
and that of edges $E(G)$; 
that is, each node represents a single spin and $E(G)$ denotes the 
set of pair of spins connected with each other. 
We make two assumptions on the system as follows. 
First, the interaction between the nodes is given by the XY 
coupling Hamiltonian \cite{Lieb1961}: 
\begin{equation}
\label{true Hamiltonian}
     H=\sum_{ (j,k) \in E(G) }
        \lambda_{jk}
         (\sigma_j^x \otimes \sigma_k^x 
            + \sigma_j^y \otimes \sigma_k^y), 
\end{equation}
where $\sigma_{j}^{x}, \sigma_{j}^{y}$, and $\sigma_{j}^{z}$ are 
the Pauli matrices acting on the $j$th spin; 
thus the notation means e.g. 
$\sigma_j^x
=I_2 \otimes \cdots \otimes \sigma^x \otimes \cdots \otimes I_2$. 
The second assumption is that only the first node is accessible; 
here we continuously measure the $z$ component of the first spin, 
in which case the measurement operator $c$ in Eqs.~\eref{eq_sme1} 
and \eref{eq_sme2} is given by 
\begin{equation}
\label{measurement operator}
    c=\sigma_{1}^{z} = \sigma^{z}\otimes I^{\otimes(N-1)}. 
\end{equation}
As a result, the conditional state of the whole spin network 
with graph $G$, whose first node is continuously measured, is 
subjected to the SME \eref{eq_sme1} with Hamiltonian 
\eref{true Hamiltonian} and the measurement operator 
\eref{measurement operator}.

%%%%%%%%%%%%%%%%%%%%%%%%%%%%%%%%%%%%%%%%%%%%%%%%%%%%%%%%%%%%%%%%%%%
%%%%%%%%%%%%%%%%%%%%%%%%%%%%  SEC.III  %%%%%%%%%%%%%%%%%%%%%%%%%%%%
%%%%%%%%%%%%%%%%%%%%%%%%%%%%%%%%%%%%%%%%%%%%%%%%%%%%%%%%%%%%%%%%%%%

\section{Structure identification via continuous measurement 
on single node}

\subsection{The structure estimator}

\begin{figure} [!t]
\begin{center}
\includegraphics[width=14cm,clip]{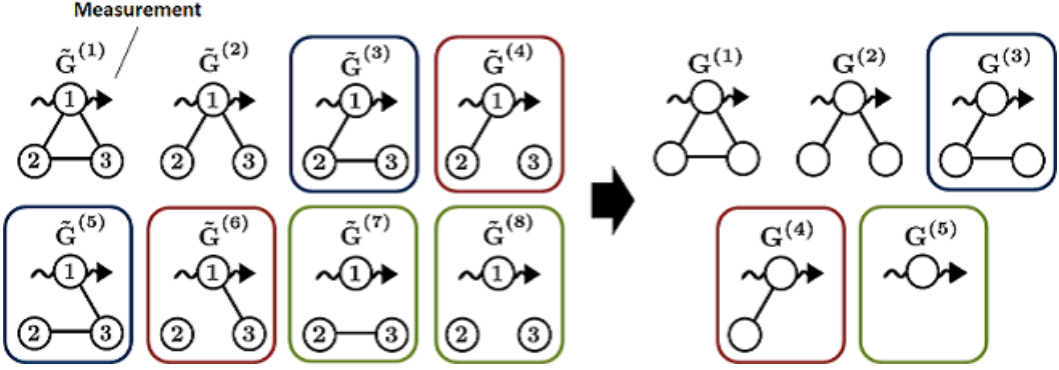}
\caption{
Possible graph structures of a three-nodes spin network. 
In each case the first node is measured as indicated by the 
wavy arrow. The graphs $\tilde{G}^{(1)}, \ldots, \tilde{G}^{(8)}$
are classified into the graphs $G^{(1)}, \ldots, G^{(5)}$, taking 
into account the topology of the graphs.}
\label{fig_2}
\end{center}
\end{figure}

We are concerned with the situation where the structure of graph 
$G$ is unknown; 
that is, we want to know which nodes of the network are connected 
with each other and how strong those connections are. 
But this general setting makes the problem too difficult, thus 
let us temporarily assume that the coupling constants $\lambda_{jk}$ 
are known and uniformly given by $\lambda$. 
Of course this assumption does not hold in general, so we will return 
to the original problem in Section 3.4. 
Also the measurement strength $\gamma$ is assumed to be known. 
Consequently, here we concentrate on the problem of identifying the 
structure of the graph $G$. 
This is equivalent to correctly choosing the {\it true graph} 
$\tilde{G}^{(i_0)}$ from all possible {\it nominal graphs} 
$\tilde{G}^{(1)}, \ldots, \tilde{G}^{(m')}$, where $m'=2^{{}_N C_2}$ 
is the number of all combinations of the edges contained in the 
$N$-spins network. 
For a network composed of three spins, for instance, we have 
totally eight candidates of graph, 
$\tilde{G}^{(1)}, \ldots, \tilde{G}^{(8)}$, whose edges are 
respectively given by 
$E(\tilde{G}^{(1)})= \{ \{1,2\},\{1,3\},\{2,3\} \}$, 
$E(\tilde{G}^{(2)})= \{ \{1,2\},\{1,3\} \}$, 
$E(\tilde{G}^{(3)})= \{ \{1,2\},\{2,3\} \}$, $\ldots$ and 
$E(\tilde{G}^{(8)})= \{ \}$, as shown in the left side of 
Fig.~\ref{fig_2}.

Clearly, the above classification is redundant, because the 
observer who accesses only to the first node cannot distinguish 
for instance $\tilde{G}^{(3)}$ and $\tilde{G}^{(5)}$; 
hence these two graphs have to be identified as the graph $G^{(3)}$, 
which is shown in the right side of Fig.~\ref{fig_2}. 
From the same reason, $\tilde{G}^{(4)}$ and $\tilde{G}^{(6)}$ 
are identified as the two-nodes graph $G^{(4)}$. 
Also $\tilde{G}^{(7)}$ and $\tilde{G}^{(8)}$ correspond to 
$G^{(5)}$. 
Consequently, for the three-nodes spin network, we can reduce 
the number of possible graph structure from $m'=8$ to $m=5$; 
the true graph $G^{(i_{0})}$ is included in the set 
${\cal G}_3=\{G^{(1)}, \ldots, G^{(5)}\}$. 
Note that ${\cal G}_3$ contains the set of single-node graph 
${\cal G}_1=\{G^{(5)}\}$ and that of two-nodes graph 
${\cal G}_2=\{G^{(4)}\}$; 
therefore, identifying the graph structure by choosing one 
element from ${\cal G}_k$ implies that at the same time we are 
estimating the number of nodes of the network, which has to be 
less than or equal to $k$ though.

To attack the problem, we employ the estimation technique, which is 
found for instance in \cite{Wiseman Book, Mabuchi 1996, Gambetta 2001,
Chase 2009, Ralph 2011, Molmer 2013}. 
The basic idea is that, based on the measurement data ${\cal Y}_t$, 
we attempt to estimate the value of both the index $i\in\{1,\ldots,m\}$ 
and a system observable in a recursive (continuous-time) manner. 
For this purpose, let us define the classical probability distribution 
$\{p^{(1)}_t, \ldots, p^{(m)}_t\}$ with 
$p_t^{(i)}={\mathbb P}(\{ G=G^{(i)} \}\hspace{0.1em}|\hspace{0.1em}{\cal Y}_t)$ 
denoting the conditional probability that the true graph of the network is 
given by $G^{(i)}$. 
Then the above-mentioned goal can be attained by constructing an update 
law of $\{p_t^{(i)}\}$ such that it changes in time and will get the maximum 
value at the index $i=i_0$. 
At the same time, we need to update the system state conditioned on 
the measurement results ${\cal Y}_t$; 
let us denote $\rho_t^{(i)}$ the whole network state corresponding to 
the $i$-th nominal graph $G^{(i)}$. 
Now, the system with graph $G^{(i)}$ is driven by the Hamiltonian 
\begin{equation}
\label{ith Hamiltonian}
    H^{(i)}
      =\sum_{ (j,k) \in E(G^{(i)}) }
        \lambda(\sigma_{j}^{x} \otimes \sigma_{k}^{x} 
          + \sigma_{j}^{y} \otimes \sigma_{k}^{y}), 
\end{equation}
while the measurement operator \eref{measurement operator} 
is commonly taken for all nominal graphs. 
By using basically the same technique for deriving the SME \eref{eq_sme1} 
and \eref{eq_sme2}, we have the following update laws of $\rho^{(i)}_{t}$ 
and $p^{(i)}_{t}$ 
(two methods to derive these equations are given in Appendix~A): 
\begin{eqnarray}
\label{eq_chase rho}
& & \hspace*{-2em}
    d \rho^{(i)}_{t} 
      = -i[H^{(i)}, \rho^{(i)}_{t}]dt 
          + \gamma \mathcal{D}[c]\rho^{(i)}_{t}dt
       + \sqrt{\gamma} \mathcal{H}[c]\rho^{(i)}_{t}
             (dY_t - 2\sqrt{\gamma}\Tr(c \rho^{(i)}_t)dt), 
\\ & & \hspace*{-2em}
\label{eq_chase p}
    dp^{(i)}_{t} 
      = 2\sqrt{\gamma} \big\{ \Tr(c \rho^{(i)}_{t})
          - \Tr(c \tilde\rho_{t}) \}p^{(i)}_{t}
              (dY_t - 2\sqrt{\gamma}\Tr(c \tilde\rho_t)dt),
\end{eqnarray}
where $\tilde\rho_{t}:=\sum_{i=1}^m p_t^{(i)}\rho_t^{(i)}$. 
Here $Y_t$ is the measurement result generated from the true 
system having the true Hamiltonian $H = H^{(i_0)}$ and the measurement 
operator \eref{measurement operator}: i.e., 
\begin{eqnarray}
\label{true sme rho}
& & \hspace*{-2em}
     d \rho^{(i_0)}_{t} 
      = -i[H^{(i_0)}, \rho^{(i_0)}_{t}]dt 
          + \gamma \mathcal{D}[c]\rho^{(i_0)}_{t}dt
       + \sqrt{\gamma} \mathcal{H}[c]\rho^{(i_0)}_{t}dW_{t}, 
\\ & & \hspace*{-2em}
\label{true sme Y}
     dY_{t} 
        = 2\sqrt{\gamma}\hspace{0.05cm}{\rm Tr}(c\rho^{(i_0)}_{t})dt 
           + dW_t.
\end{eqnarray}
We recursively calculate the above equations to update the probability 
distribution $p_t^{(i)}$ as well as the state $\rho_t^{(i)}$, using 
the measurement result $Y_t$; 
what we expect is that, again, $p_t^{(i)}$ will get the maximum value at 
the index $i=i_0$ after many iterations. 
Note that, in reality Eqs.~\eref{true sme rho} and \eref{true sme Y} 
cannot be computed since $H^{(i_0)}$ is unknown, but only the experimental 
data $Y_t$ is obtained; 
in numerical simulations, however, we do that in order to generate $Y_t$. 
Figure 3 illustrates the configuration of the estimation scheme.

\begin{figure} [!t]
\begin{center}
\includegraphics[width=12.5cm,clip]{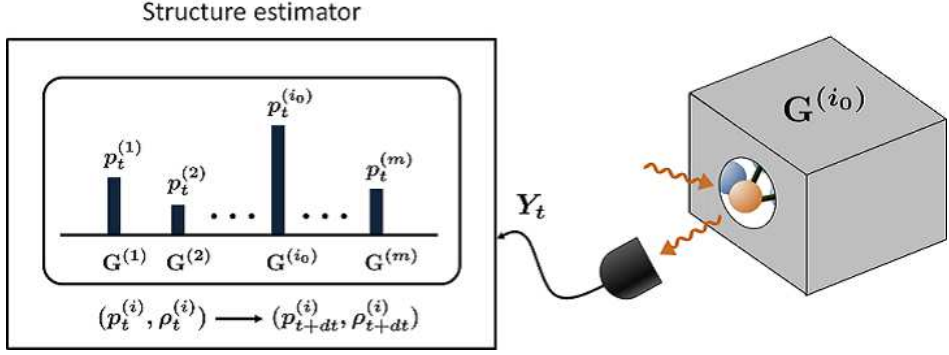}
\caption{
Configuration of the structure estimator. 
We perform a continuous-time measurement on the accessible node of the 
spin network whose graph structure $G^{(i_0)}$ is unknown. 
The measurement result $Y_t$ is used to update $p_t^{(i)}$, the probability 
that $G^{(i)}$ is the true graph, as well as $\rho_t^{(i)}$, the quantum state 
of the system with graph $G^{(i)}$. 
The update laws are given by Eqs.~\eref{eq_chase rho} and 
\eref{eq_chase p}. 
}
\label{fig_3}
\end{center}
\end{figure}
%

%%%%%%%%%%%%%%%%%%%%%%%%%%%%%%%%%%%%%%%%%%%%%%%%%%%%%%%%%%%%%%%%%%%

\subsection{Example 1: three-spins case}

\begin{figure} [!t]
\begin{center}
\includegraphics[width=15.5cm,clip]{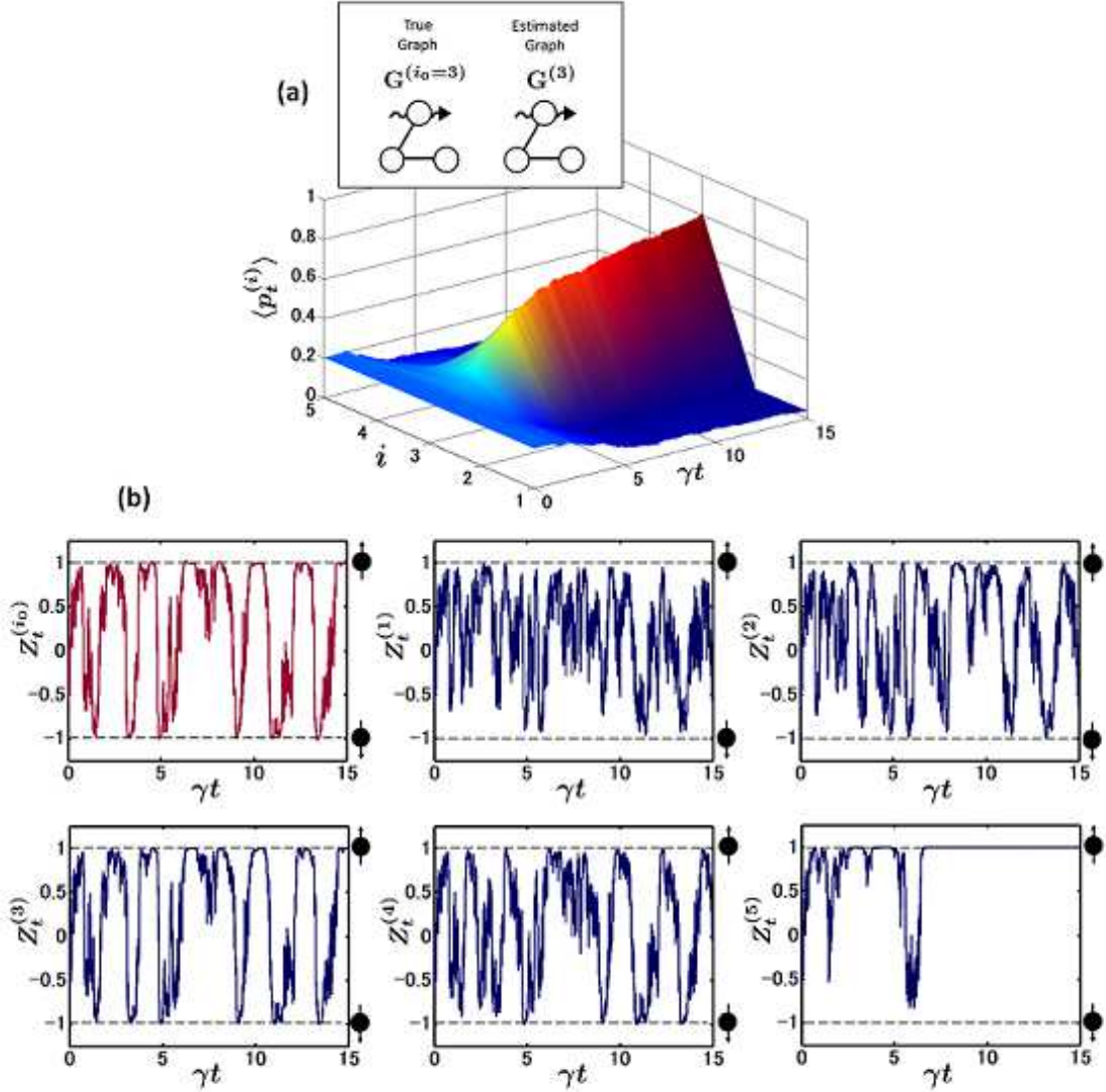}
\caption
{
Numerical simulation for the three-spins network with parameter 
$\lambda = \gamma$. 
The true graph structure is set to $G^{(i_0=3)}$. 
Figure (a) shows the time evolution of the probability distribution 
$\mean{ p^{(i)}_t }$ averaged over 50 sample paths, which takes 
the highest value at the true index $i_0=3$. 
Figure (b) shows typical sample paths of the estimate of $c=\sigma_1^z$, 
for the true system with graph $G^{(i_0=3)}$ (upper left) and for the 
nominal systems. 
}
\label{fig_4}
\end{center}
\end{figure}

Let us consider the simple network composed of three nodes; 
in this case, as depicted in Fig.~\ref{fig_2}, we have $m=5$ 
candidates as the graph structure. 
The true system is chosen to be the chain-type network $G^{(i_0=3)}$. 
The initial distribution is set to the uniform one $p^{(i)}_0=1/5~\forall i$, 
because the graph structure is assumed to be completely unknown at 
the initial time $t=0$. 
From a similar reason, we should set the initial density matrix 
to the maximal mixed state $\rho^{(i)}_0=(I_2/2)^{\otimes 3}~\forall i$. 
In this setting, we run the algorithm \eref{eq_chase rho} and 
\eref{eq_chase p} to compute $p^{(i)}_t$. 
Figure~4~(a) shows the averaged time evolution of 50 sample paths of 
$p^{(i)}_t$, denoted by $\mean{p_t^{(i)}}$; 
from this we clearly see that the correct convergence of $p^{(i)}_t$ to the 
distribution with $p^{(3)}=1$ occurs most frequently. 
Hence, our estimator correctly identifies the true graph $G^{(3)}$.

We now discuss why the identification is possible by measuring only 
a part of the network. 
For this purpose let us focus on the estimate (conditional 
expectation) of the $z$-component of the measured spin. 
The continuous measurement tends to increase the absolute value of 
the estimate of $\sigma^z$ \cite{Handel 2005}, while now the value 
of the $z$-component of the measured spin is distributed over the 
network due to the XY coupling Hamiltonian \cite{Bloembergen,Negoro}; 
i.e. {\it spin diffusion} occurs. 
Hence, intuitively, if the network is ``small" in the sense that 
the path length from the accessible node to every terminal nodes 
is relatively short, then the spin wave quickly gets back to the 
measured spin and consequently the estimate of the $z$-component 
of the measured spin will change very fast, while in the opposite 
case the estimate will change slowly. 
Figure~\ref{fig_4}~(b) plots the trajectories of 
$Z^{(i_0=3)}_{t}=\Tr(c \rho^{(i_0=3)}_{t})$ and 
$Z^{(i)}_t=\Tr(c \rho^{(i)}_t)$. 
These figures support the validity of the above observation; 
because the chain is relatively a ``large" network, the true 
estimate $Z^{(i_0=3)}_t$ actually changes slowly. 
Remarkably, only the nominal estimate $Z^{(3)}_t$ shows a similar 
trajectory to that of the true one $Z_t^{(i_0=3)}$, while the 
other nominals do not. 
This fact means that the measurement even only on a part of the 
network certainly brings useful information for identifying 
the whole structure. 
At the same time, Fig.~\ref{fig_4}~(b) tells us that the time-evolution 
of $Z^{(3)}_t$ produced from the large network is singularly different 
from those produced from the small networks, i.e. $Z^{(1)}_t$, 
$Z^{(2)}_t$, and $Z^{(4)}_t$, which all behave in a similar fashion. 
In general, if the interaction strength are uniform and the upper bound 
of total spin number of the network is known, then there are a few large 
systems having similar graphs, while there may be many small systems 
with similar structure; 
thus, it is expected that a large network tends to produce a singular signal 
that allows us to easily distinguish it from others, while not the case for 
a small network.

%%%%%%%%%%%%%%%%%%%%%%%%%%%%%%%%%%%%%%%%%%%%%%%%%%%%%%%%%%%%%%%%%%%

\subsection{Example 2: five-spins case}

\begin{figure} [!t]
\begin{center}
\includegraphics[width=15.5cm,clip]{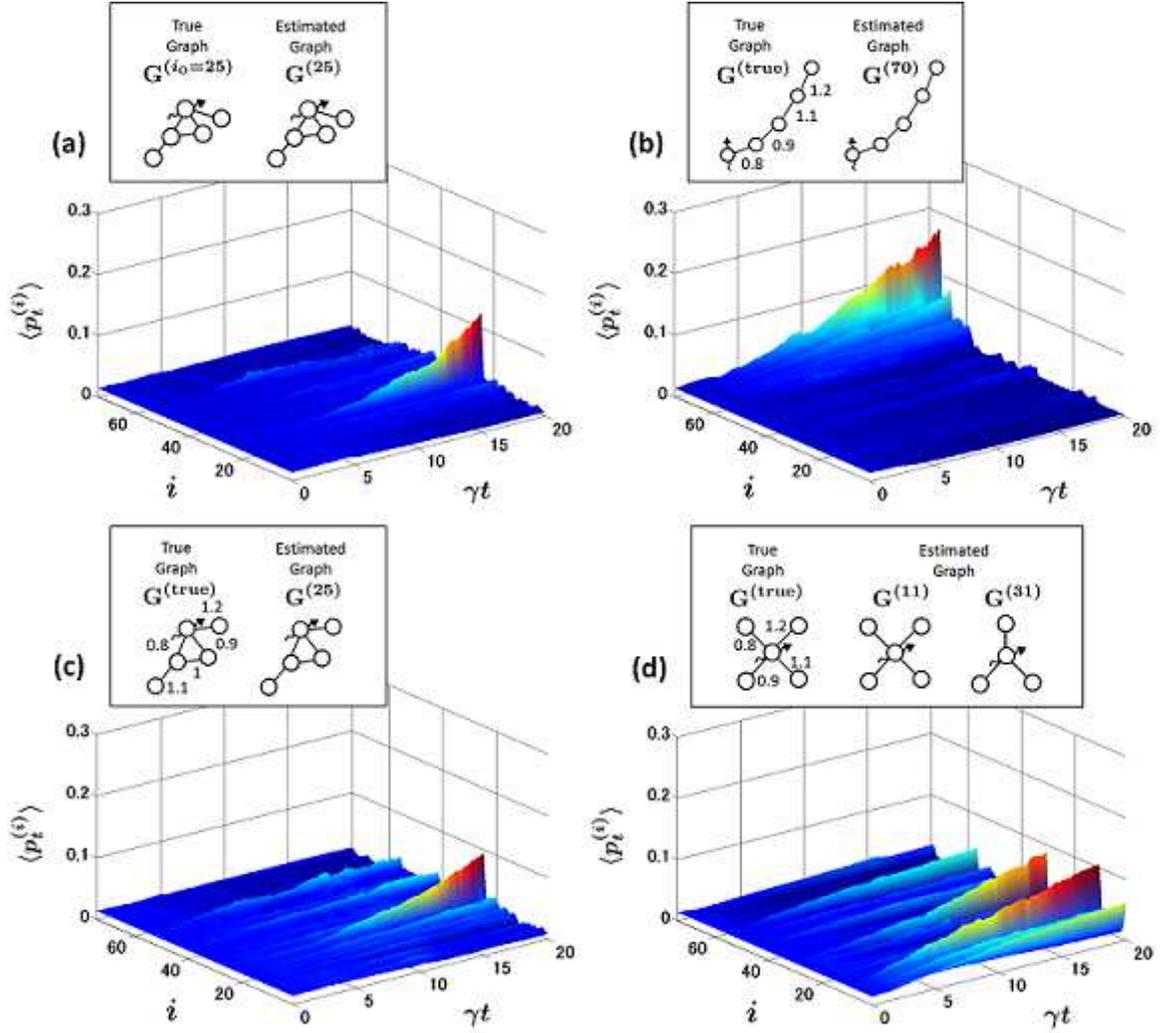}	
\caption{
(a) 
Time evolution of the probability distribution $\mean{ p^{(i)}_t }$ 
averaged over 50 sample paths. 
The true graph is $G^{(i_0=25)}$, which has the uniform coupling 
constant. 
The parameters are set to $\lambda = \gamma$. 
(b,c,d) 
Several examples where the coupling constants of the true network 
are not uniform. 
The number along the edge $\{j,k\}$ represents 
$\lambda_{jk}/\gamma$. 
The time evolution of $\mean{p^{(i)}_t}$ is computed by averaging 
50 sample paths. 
The true graph and the estimated graph are both depicted in the inset. 
}
\label{fig_5}
\end{center}
\end{figure}

We next consider the five-nodes network with true graph $G^{(i_0=25)}$, 
which is depicted in the inset of Fig.~\ref{fig_5} (a). 
For networks composed of up to five nodes, there are totally $m=74$ 
graph structures, so the true graph is contained in the set 
${\cal G}_5=\{G^{(1)}, \ldots, G^{(74)}\}$. 
Note that, without taking into account the redundancy, the number 
of possible graph structure is $m'=1024$; 
hence the efficiency of the classification method introduced in 
Sec.~3.1 warrants special mention.

The time evolution of $\mean{p_t^{(i)}}$ is computed by averaging 
over 50 sample paths of $p_t^{(i)}$ and shown in Fig.~\ref{fig_5}~(a). 
The initial distribution is set to the uniform one 
$p^{(i)}_0=1/74~\forall i$, because of the same reason explained 
before. 
Also $\rho^{(i)}_0=(I_2/2)^{\otimes 5}~\forall i$. 
Note that this system is a relatively small network with length of 
up to 2 from the accessible node; 
hence, as discussed in Sec.~3.2, the system may be less distinguishable 
compared to the chain-structured one. 
Nonetheless, the trajectory still converges and takes the maximum 
value at $i=25$, thus the estimator correctly identifies the true 
graph structure.

%

%%%%%%%%%%%%%%%%%%%%%%%%%%%%%%%%%%%%%%%%%%%%%%%%%%%%%%%%%%%%%%%%%%%

\subsection{Example 3: Networks with non-uniform coupling constants}

We have observed that our estimator can identify the true graph 
structure with high probability, under the assumption that the 
coupling constants of the true network are known and uniform. 
Thus here we should return to the original problem, i.e. 
structure identification of a network having {\it non-uniform} 
coupling constants. 
So the true network has the Hamiltonian of the form 
\begin{eqnarray}
\label{non-uniform Hamiltonian}
& & \hspace*{-1em}
     H^{({\rm true})} = \sum_{ (j,k) \in E(G^{({\rm true})}) }\lambda_{jk}
         (\sigma_j^x \otimes \sigma_k^x 
              + \sigma_j^y \otimes \sigma_k^y). 
\end{eqnarray}
$G^{({\rm true})}$ is the true graph with unknown coupling constants 
$\lambda_{jk}$. 
To attack this general problem, in this paper we follow the strategy 
to estimate only the graph structure; 
that is, we apply the same estimator as before, which assumes the 
{\it uniform} coupling constants for the underlying network. 
Thus the estimator attempts to choose a most-likely graph from the 
set ${\cal G}_m=\{G^{(1)},\ldots,G^{(m)}\}$ composed of the graphs 
with uniform coupling constants; 
hence $G^{({\rm true})}$ is not contained in ${\cal G}_m$. 
There are two related reasons behind this approach; 
first, as mentioned in Sec.~1, if the graph structure is correctly identified, 
then the coupling strength can be estimated using the method developed 
in \cite{Burgarth 2009 1, Burgarth 2009 2,Franco 2009,Burgarth 2011, 
Lapasar 2012, Maruyama 2012}; 
second, it has a clear advantage in computational time, because identifying 
both the graph structure and coupling constants in the same time means 
that we update $q_t^{(i,j)}$, the probability distribution of the $j$-th 
coupling constant of $G^{(i)}$, in addition to $(p_t^{(i)}, \rho_t^{(i)})$, 
which would be numerically intractable even for a small-size system.

%Note again that the estimator is driven by the measurement result 
%obtained at the accessible node, $Y_t$, which is generated from the 
%true network with the weighted graph $G^{({\rm true})}$. 

Here, we consider three five-nodes spin networks. 
The first one is the case where the true network is of the chain 
structure depicted in the inset of Fig.~\ref{fig_5}~(b). 
Since a chain is the largest network, as seen in Fig.~\ref{fig_4}~(b), 
the estimate of the $z$-component of the measured spin should be 
singularly slow in changing, implying that the chain-type network 
may be easily distinguished from other candidates even in the 
non-uniform case. 
In fact, Fig.~\ref{fig_5}~(b) shows that the estimator correctly 
identifies the true graph structure $G^{(70)}$.

The next is the case where the true graph $G^{({\rm true})}$ is 
shown in the inset of Fig.~\ref{fig_5}~(c). 
As discussed in Sec.~3.3, this is a relatively small network, hence 
the distinguishability, or roughly speaking the identifiability, 
would become worse. 
Figure~\ref{fig_5}~(c) shows this is indeed the case; 
although $\mean{p^{(i)}_t}$ takes the maximum values at $i=25$ and 
thus the estimator correctly identifies the true graph structure, 
the probability to reach the point $i=25$ becomes smaller than the 
case of chain. 
Note also that the success probability becomes smaller than the 
previous case shown in Fig.~\ref{fig_5}~(a).

Lastly, we consider the star-type network with the weighted graph 
$G^{({\rm true})}$ depicted in the inset of Fig.~\ref{fig_5}~(d). 
In this case, as shown in the figure, $\mean{p^{(i)}_t}$ has 
two comparable peaks at the points corresponding to the correct 
answer $G^{(11)}$ and the wrong one $G^{(31)}$; 
although $\mean{p^{(11)}_t}$ is a bit larger than $\mean{p^{(31)}_t}$, 
we should take both answers as the identified model. 
That is, the estimator cannot definitively identify the true graph. 
This result makes sense, because the true system is a smallest 
five-nodes network that would have a number of other types of 
small networks producing a similar output signal $Y_t$; 
in particular, $G^{(11)}$ and $G^{(31)}$ generate very similar time evolutions 
of the estimate of $c=\sigma_1^z$, hence it is hard to distinguish them.

%%%%%%%%%%%%%%%%%%%%%%%%%%%%%%%%%%%%%%%%%%%%%%%%%%%%%%%%%%%%%%%%%%%

\subsection{Discussion}

Our main question is the following: 
for what kind of network systems does the algorithm work well and 
generate the correct answer? 
The key concept enabling us to approach this problem would be 
{\it identifiability}, which has already appeared above without 
a formal definition: 
That is, if two different systems generate different outputs for 
a given common input, they can be distinguished from the output data 
and thus called identifiable. 
This is a fundamental notion in the field of {\it system 
identification} \cite{LjungBook}, whose quantum version has started 
to be studied only very recently \cite{Burgarth PRL 2012,
Guta Yamamoto 2012}. 
So our conjecture is as follows.

{\bf Conjecture 1:} 
If the true system is identifiable, then the solution of the SME converges 
to the correct answer with high probability.

Actually, it can be proven that the chain-formed network is 
identifiable \cite{Burgarth PRL 2012,Guta Yamamoto 2012}. 
On the other hand the system in Fig.~5~(d) seems to be not 
identifiable, which can be seen from the fact that the systems 
$G^{(11)}$ and $G^{(31)}$ generate almost the same amount 
of the probability $\mean{p^{(i)}_t}$. 
Exploring the connection of the structure identification problem 
to the general identifiability analysis is very important and 
should bring useful facts.

%For instance, it is known in classical case that the identifiability 
%property generally depends on the injected input; 
%hence using the single photon filter \cite{Gough} would make the 
%system in Fig.~4~(d) even to generate different output and consequently 
%give us a chance to correctly identify network structure. 

%%%%%%%%%%%%%%%%%%%%%%%%%%%%%%%%%%%%%%%%%%%%%%%%%%%%%%%%%%%%%%%%%%%
%%%%%%%%%%%%%%%%%%%%%%%%%%%%%  SEC.IV  %%%%%%%%%%%%%%%%%%%%%%%%%%%%
%%%%%%%%%%%%%%%%%%%%%%%%%%%%%%%%%%%%%%%%%%%%%%%%%%%%%%%%%%%%%%%%%%%

\section{State initialization via continuous measurement 
on single node}

If the network structure is correctly identified, the estimation 
methods developed in \cite{Burgarth 2009 1,Burgarth 2009 2,
Franco 2009,Burgarth 2011, Lapasar 2012, Maruyama 2012} can be 
applied to determine the coupling constant $\lambda_{jk}$ in the 
Hamiltonian \eref{true Hamiltonian}, and then we can further move 
forward to the stage of system initialization. 
In this section, under the assumption that the Hamiltonian 
\eref{true Hamiltonian} is completely known, we provide a scheme 
that deterministically stabilizes the spin coherent state 
$\ket{0^{\otimes N}}=\ket{0, 0, \cdots, 0}$. 
The scheme is again based on the continuous measurement performed 
only on a single node of the network. 
As mentioned in Sec.~1, in general, measurement can reduce the 
entropy of the system state, while it must drive the state 
probabilistically; 
to achieve the deterministic state preparation, we thus employ 
the mechanism of adaptive measurement. 
To make this idea clear, this section is first devoted to present an 
adaptive measurement method for state preparation of a single-spin 
system. 
Then it is applied to quantum networks with only a single node 
accessible.

Before describing the results, we make two remarks. 
First, for a solid system, we can effectively prepare a product of 
ground states $\ket{g, g, \cdots, g}$ by cooling the system. 
The situation considered in this paper, however, does not allow 
the standard cooling method that extracts entropy from all the 
nodes via the global system-refrigerator coupling. 
Rather we here need to extract entropy from only a part of the system. 
Second, a completely different type of initialization method was 
proposed in \cite{Burgarth 2007}; 
the idea is to utilize a state transfer architecture from an ancilla 
system to the network system through accessible nodes. 
This method is effective if initializing the ancilla can be 
done easily.

%%%%%%%%%%%%%%%%%%%%%%%%%%%%%%%%%%%%%%%%%%%%%%%%%%%%%%%%%%%%%%%%%%%

\subsection{Adaptive measurement for single spin state preparation}

As indicated by Eq.~\eref{eq_sme1}, measuring a quantum system 
always brings a stochastic driving of the state. 
This means that only a fixed measurement does not deterministically 
stabilize the state. 
Combining measurement with feedback control is thus expected to 
overcome this issue, as actually demonstrated in several studies 
\cite{Wiseman Book,Stockton 2004,Handel 2005,Yamamoto 2007,Mirrahimi}. 
Adaptive measurement is a kind of feedback control, which does 
not introduce an additional actuator for control but instead changes 
the detector configuration based on the past measurement results. 
Thus a merit of adaptive measurement may appear in a practical 
situation where it costs cheaper than adding an additional actuator. 
The applicability of this scheme to single spin stabilization 
has been demonstrated in \cite{Jacobs 2010,Tanaka 2012}; 
we here consider the same problem studied in these references 
and show a new result.

The problem of stabilizing a single spin state via adaptive 
measurement is described as follows. 
Recall that the state under continuous measurement evolves in time 
according to Eq.~\eref{eq_sme1}. 
We here set $H=0$ and assume that the measurement operator $c$ 
can be changed in time as a function of $\rho_t$; 
let us parameterize $c$ in the following form:
\begin{eqnarray}
\label{adaptive c}
   c_t= \left(\begin{array}{cc}
         \cos \theta_t & e^{-i \delta_t} \sin \theta_t  \\
         e^{i \delta_t} \sin \theta_t &  -\cos \theta_t \\
      \end{array}\right). 
\end{eqnarray}
Then the problem is to determine the time evolution of the 
parameters $(\theta_t, \delta_t)$ so that the single 
spin state $\rho_t$ governed by the SME \eref{eq_sme1} with 
$H=0$ and Eq.~\eref{adaptive c} deterministically converges 
to a desired target state. 
In particular, we set the target to be $\ket{0}$.

To solve the problem, let us consider the following cost function: 
\begin{equation*}
    J_t=1-\Tr(\sigma^z \rho_t). 
\end{equation*}
This is non-negative and takes the minimum value $0$ only when 
$\rho_t= \ket{0}\bra{0}$. 
Also we parameterize the state as 
\begin{equation}
\label{state in the proof}
   \rho_t 
      = \frac{1}{2}
        \left( \begin{array}{cc}
          1+r_t\cos\alpha_t & r_t e^{-i \beta_t} \sin\alpha_t  \\
          r_t e^{i \beta_t} \sin \alpha_t  & 1-r_t\cos\alpha_t \\
        \end{array} \right). 
\end{equation}
Then, the derivative of ${\mathsf E}[J_t]$ is given by 
\begin{equation}
\label{eq_de1}
   \frac{d{\mathsf E}[J_t]}{dt} 
      = -\frac{r_t}{2} \Big[
          (\cos(2\theta_t) - 1) \cos\alpha_t
        + \sin(2\theta_t) \sin\alpha_t \cos(\delta_t -\beta_t) \Big].
\end{equation}
Hence, by choosing the tuning parameters as 
\begin{equation}
\label{adaptive law}
    (\theta_t, \delta_t)
      =(\alpha_{t}/2, \beta_{t}),~~(-\alpha_{t}/2, \beta_{t}+\pi), 
\end{equation}
we have 
\[
    \frac{d{\mathsf E}[J_t]}{dt}
      =-\frac{r_t}{2}(1-\cos\alpha_t) \leq 0. 
\]
Then, from the theory of stochastic stability \cite{Hasminskii}, 
$d{\mathsf E}[J_t]/dt\rightarrow 0$ holds, thus equivalently 
$\alpha_t\rightarrow 0$ or $r_t\rightarrow 0$ is guaranteed. 
This means that after long time limit the state lies on the positive 
half of the $z$ axis in the Bloch sphere. 
But it is well known that an ideal continuous measurement of an 
observable always increases the purity of the conditional state; 
i.e., $r_t\rightarrow 1$. 
Combining these two results, we can conclude that the state 
converges to the target state $\ket{0}$ almost surely. 
The adaptive measurement law \eref{adaptive law} has been found 
in \cite{Tanaka 2012}, though without rigorous proof. 
Hence here we present the result as a new contribution.

{\bf Theorem 1}: 
The single spin state subjected to the SME \eref{eq_sme1} with 
$H=0$ and the adaptive measurement law \eref{adaptive c}, 
\eref{state in the proof}, and \eref{adaptive law} converges 
to the target state $\ket{0}$ almost surely.

%%%%%%%%%%%%%%%%%%%%%%%%%%%%%%%%%%%%%%%%%%%%%%%%%%%%%%%%%%%%%%%%%%%

\subsection{Network initialization via adaptive measurement 
on single node}

Now we apply the adaptive measurement scheme developed in the 
previous subsection to an $N$-spins quantum network with only a 
single node accessible. 
Let us set the target to be the spin coherent state 
$\ket{0^{\otimes N}}$. 
Then, the goal is to stabilize the target $\ket{0^{\otimes N}}$ by 
applying the continuous adaptive measurement performed on the 
accessible spin.

We follow the same procedure as before. 
Now the whole state is governed by the SME \eref{eq_sme1} with 
Hamiltonian \eref{true Hamiltonian} and the measurement operator 
\begin{equation}
\label{network adaptive}
     c'_t = \left(\begin{array}{cc}
         \cos\theta'_t & e^{-i \delta'_t} \sin\theta'_t \\
         e^{i \delta'_t}\sin\theta'_t & -\cos\theta'_t  \\
       \end{array}\right)
          \otimes I^{\otimes(N-1)}, 
\end{equation}
instead of Eq.~\eref{measurement operator}. 
The adaptive measurement law of the parameters $(\theta'_t, \delta'_t)$ 
can be determined from the following cost function: 
\begin{eqnarray}
     J'_t = N - \Tr\Big( \sum_{j=1}^{N}\sigma_j^z \rho_t \Big),
\end{eqnarray}
which is non-negative and takes the minimum value $0$ only when 
$\rho_t=\ket{0^{\otimes N}}\bra{0^{\otimes N}}$. 
The time derivative of ${\mathsf E}[J'_t]$ is given by 
\begin{equation}
\label{eq_de2}
   \frac{d{\mathsf E}[J'_t]}{dt} 
      = -\frac{r'_t}{2} \Big[
          (\cos(2\theta'_t) - 1) \cos\alpha'_t
        + \sin(2\theta'_t) \sin\alpha'_t 
              \cos(\delta'_t -\beta'_t) \Big], 
\end{equation}
where $(r'_t,\alpha'_t,\beta'_t)$ are the parameters of the 
reduced quantum state 
\begin{equation}
   \rho'_t 
      = \Tr_{(2,3,4,...,N)}[\rho_t]
      = \frac{1}{2}
        \left( \begin{array}{cc}
          1+r'_t\cos\alpha'_t & r'_t e^{-i \beta'_t} \sin\alpha'_t \\
          r'_t e^{i \beta'_t} \sin \alpha'_t & 1-r'_t\cos\alpha'_t \\
        \end{array} \right). 
\end{equation}
The point is that Eq.~\eref{eq_de2} has the same form as that 
for the single spin case, Eq.~\eref{eq_de1}. 
Thus the adaptive law 
\begin{equation}
\label{adaptive law network}
    (\theta'_t, \delta'_t)
      =(\alpha'_t/2, \beta'_t),~~(-\alpha'_t/2,\beta'_t+\pi)
\end{equation}
gives rise to $d{\mathsf E}[J'_t]/dt\rightarrow 0$ as before, 
which thus concludes $\rho'_t\rightarrow \ket{0}\bra{0}$. 
Also note that the measurement operator \eref{network adaptive} 
then converges to Eq.~\eref{measurement operator}.

Although the above result does not necessarily mean the deterministic 
convergence of the whole network state $\rho_t$ to the target 
$\ket{0^{\otimes N}}$, the following two facts suggest that it would 
actually occur. 
First, we now know that the value of the $z$-component of the first 
spin, which is continuously raised via the adaptive measurement, 
is distributed over the whole network due to the XY coupling 
Hamiltonian; 
hence the $z$-components of all spins may also increase. 
Second, it can be proven that the target $\ket{0^{\otimes N}}$ 
is a steady state of the SME with Eqs.~\eref{true Hamiltonian} and 
\eref{measurement operator} (see Appendix~B). 
In view of these two facts, we pose the following conjecture.

{\bf Conjecture 2:} 
The whole network state $\rho_t$ will deterministically converge 
to the target $\ket{0^{\otimes N}}$, if it is the {\it unique} 
steady state of the {\it controlled SME}, i.e. the SME containing 
the adaptive measurement schematic.

%%%%%%%%%%%%%%%%%%%%%%%%%%%%%%%%%%%%%%%%%%%%%%%%%%%%%%%%%%%%%%%%%%%

\subsection{Example: five-spins network}

\begin{figure} [!t]
\begin{center}
\includegraphics[width=16cm]{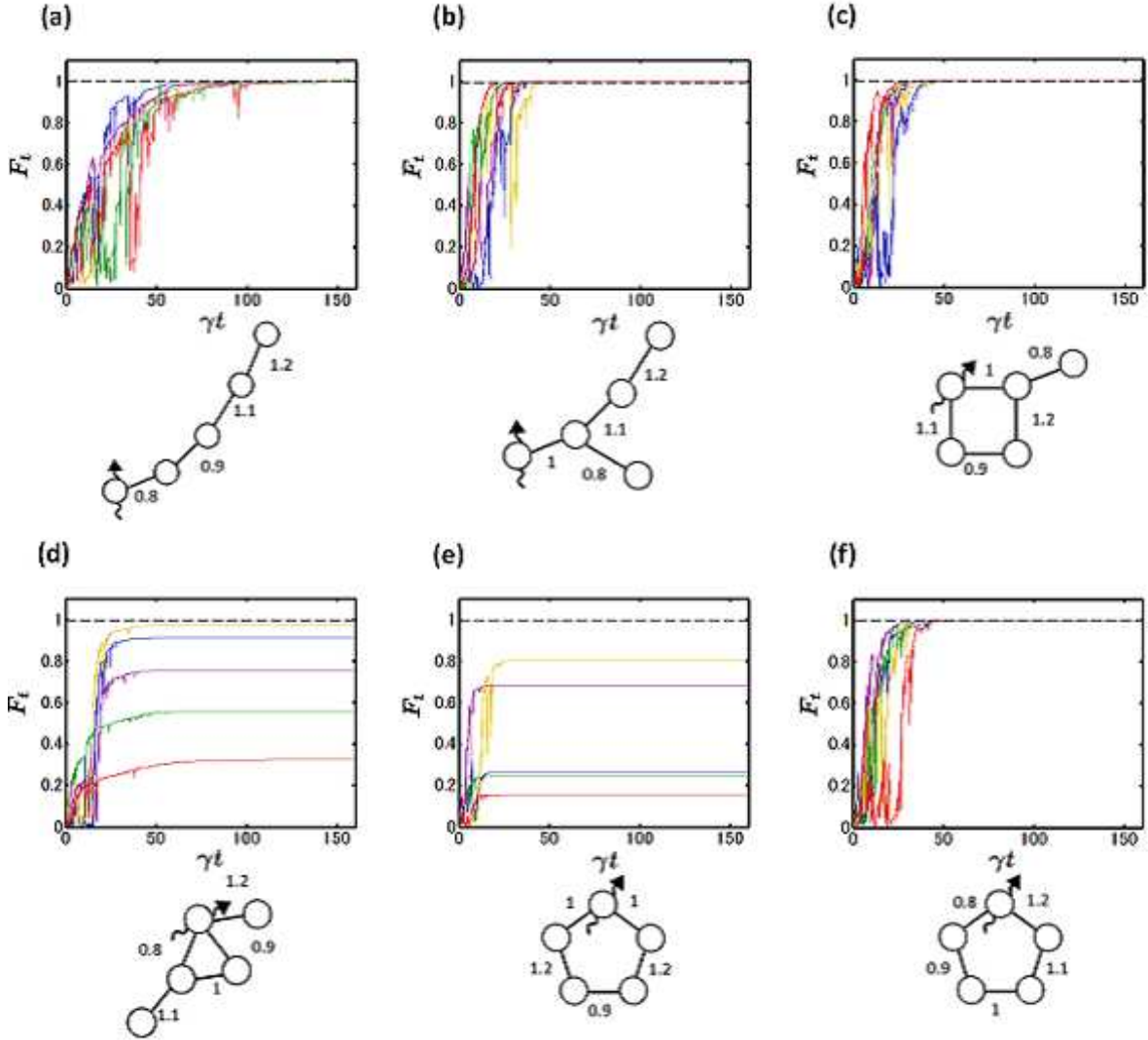}
\caption{
Sample paths of the fidelity 
$F_t=\bra{0^{\otimes 5}}\rho_t\ket{0^{\otimes 5}}$ 
for several five-spins networks. 
The graph (e) contains the permutation symmetry structure. 
The number along the edge $\{j,k\}$ represents $\lambda_{jk}/\gamma$. 
}
\label{fig_6}
\end{center}
\end{figure}

Here we consider some five-spins networks shown in Fig.~\ref{fig_6}, 
whose first spin is continuously measured with the adaptive law 
\eref{adaptive law network}. 
To evaluate the performance, we use the fidelity 
$F_t=\bra{0^{\otimes 5}}\rho_t\ket{0^{\otimes 5}}$, which takes 
the maximum value $1$ only when 
$\rho_t=\ket{0^{\otimes 5}}\bra{0^{\otimes 5}}$. 
The initial state is $\rho_0=(I_2/2)^{\otimes 5}$. 
In each panel of Fig.~\ref{fig_6}, some sample paths of $F_t$ are 
displayed.

First let us focus on the systems shown in the figures (a,b,c,f). 
The remarkable fact is that, for all these cases, the target 
$\ket{0^{\otimes 5}}$ is the unique steady state of the controlled 
SME. 
Hence, Conjecture~2 mentioned above suggests that our scheme 
realizes the deterministic state stabilization; 
actually all trajectories converge to $1$. 
Also this result shows that, for a variety of network structure, 
the goal can be achieved.

On the other hand, the figures show that, in the cases (d) and (e), 
the adaptive scheme does not work well. 
Indeed, in these cases, it can be proven that the controlled SME 
has a steady state other than the target, implying that the 
contraposition of Conjecture~2 is true. 
In particular, the system shown in the figure (e) has a special 
symmetric structure; 
indeed this structure is what prohibits the desirable convergence 
and will be examined in detail in the next subsection.

%%%%%%%%%%%%%%%%%%%%%%%%%%%%%%%%%%%%%%%%%%%%%%%%%%%%%%%%%%%%%%%%%%%

\subsection{Permutation symmetry}

The above results imply that, as suggested by Conjecture~2, the deterministic 
convergence needs the condition that the target $\ket{0^{\otimes 5}}$ is the 
unique steady state of the controlled SME. 
Hence, it should be useful to characterize a system that does 
{\it not} have such uniqueness property. 
In particular, systems having {\it permutation symmetry} 
\cite{Wang 2012} are important; 
this property means that the Hamiltonian is invariant under the 
exchange of some specific pairs of spins. 
The system shown in Fig.~\ref{fig_6}~(e) has this property; 
actually the Hamiltonian is 
\begin{eqnarray}
& & \hspace*{-4em}
    H=  \gamma( XXIII + YYIII ) + 1.2\gamma(IXXII + IYYII) 
      + 0.9\gamma(IIXXI + IIYYI)
\nonumber \\ & & \hspace*{2em}
    \mbox{}
      + 1.2\gamma(IIIXX + IIIYY) + \gamma(XIIIX + YIIIY),
\nonumber
\end{eqnarray}
where e.g. 
$XXIII=\sigma^x\otimes\sigma^x\otimes I_2\otimes I_2\otimes I_2$, 
and it is permutation symmetric with respect to the exchanges between 
the second and the fifth spins and between the third and the fourth 
spins. 
We can then find that the entangled state 
\[
\hspace*{-4em}
     \ket{\phi'} 
        = \frac{1}{\sqrt{2(1+a^{2})}}
            \Big( \ket{00001} + a \ket{00010} 
               - a \ket{00100} - \ket{01000}\Big),~~~
     a = \frac{-3 \pm \sqrt{73}}{8}
\]
satisfies $c\ket{\phi'}=\ket{\phi'}$ and $H\ket{\phi'}=1.2a\ket{\phi'}$, 
which thus implies that, in addition to $\ket{0^{\otimes 5}}$, 
$\ket{\phi'}$ is a steady state of the controlled SME because of 
the fact described in Appendix~B. 
Note that $\ket{\phi'}$ is invariant under the above-mentioned 
permutation operation. 
Thus, the state can also move toward $\ket{\phi'}$, implying that the 
deterministic convergence to the target $\ket{0^{\otimes 5}}$ would 
not be expected. 
Actually, as seen in Fig.~\ref{fig_6}~(e), the state does not 
converge to the target, even with the aid of the adaptive measurement; 
rather it converges to a mixed state on the subspace spanned by 
the steady states of the controlled SME.

The above result can be generalized, as shown below. 
The proof is given in Appendix~C.

{\bf Theorem 2}: 
Suppose that the network has the permutation symmetry property. 
Then, the target state is not the unique steady state of the 
controlled SME.

But note again that the system shown in Fig.~\ref{fig_6}~(f) shows 
the deterministic convergence. 
Hence the deterministic stabilization of the state at the target 
is simply recovered, if the system experiences some perturbation 
and loses the permutation symmetry.

%%%%%%%%%%%%%%%%%%%%%%%%%%%%%%%%%%%%%%%%%%%%%%%%%%%%%%%%%%%%%%%%%%%

\subsection{Discussion}

The numerical simulations support the validity of Conjecture~2 
posed at the end of Section 4.2; 
if the target is the unique steady state of the controlled SME, 
then the adaptive scheme achieves the deterministic state 
initialization. 
Therefore, to make the presented scheme stronger, we need to 
prove this conjecture and also characterize the network structure 
such that the spin coherent state is the unique steady solution of 
the controlled SME. 
These problems are both difficult due to the huge variety of the 
network structure, but we expect that the results \cite{Kraus} and 
\cite{Mirrahimi} could be applied to solve them; 
the former shows the uniqueness condition of the steady state of a 
general master equation, and the latter gives a rigorous proof of 
the deterministic convergence in the case $c=\sum_k\sigma_k^z$ and 
$H=u(t)\sum_k\sigma_k^y$ with $u(t)$ the feedback input. 
Both problems need careful mathematical analysis and should be 
investigated in the future work.

%%%%%%%%%%%%%%%%%%%%%%%%%%%%%%%%%%%%%%%%%%%%%%%%%%%%%%%%%%%%%%%%%%%
%%%%%%%%%%%%%%%%%%%%%%%%%%  Conclusion  %%%%%%%%%%%%%%%%%%%%%%%%%%%
%%%%%%%%%%%%%%%%%%%%%%%%%%%%%%%%%%%%%%%%%%%%%%%%%%%%%%%%%%%%%%%%%%%

\section{Concluding discussion}

In this paper, we have provided continuous-measurement-based methods 
to achieve the structure identification and deterministic 
state initialization of spin networks with single node only 
accessible. 
In each case, as numerically demonstrated, the performance of the 
scheme fully depends on the network structure. 
So surely it is very important to clarify what kind of graph structure 
is suitable for achieving {\it both} goals. 
This general question is of course not straightforward to answer, 
but here we try to deduce a conjecture from an intuitive observation.

First, to succeed in the structure identification, we need enough 
information that can be extracted from the accessible node; 
in general, when the system is in a highly mixed state, such 
information leaking occurs. 
On the other hand, if the system state is in a pure state, or in 
our case the state initialization has been completed, no meaningful 
information is available anymore. 
Hence, it seems that the identification and the initialization 
are in a trade-off relationship. 
Is there a system that allows us to achieve both goals? 
Indeed, we have seen that the presented two schemes work well 
particularly for the chain-formed system, as observed in 
Figs.~\ref{fig_5}~(b) and \ref{fig_6}~(a). 
This can be understood by looking at the fact that the chain is 
relatively a large network that needs longer time until being 
purified; 
hence a large network offers more information during the measurement 
process compared to some other small networks. 
Indeed Fig.~\ref{fig_6}~(a) shows that the chain-formed network 
takes the longest time to be initialized. 
Based on these observation, we now have the following general conjecture.

{\bf Conjecture 3:} 
A system having the {\it infection property} 
\cite{Burgarth 2009 2,Burgarth 2011, Maruyama 2012,Burgarth 2009 3, 
Burgarth PRL 2012,Guta Yamamoto 2012} is the best suitable for 
quantum computation on a limited-access network.

Actually, an infective system is essentially equivalent to a 
chain-formed system. 
The importance of this class of systems lies in the following three facts; 
first, an infective system can be parameter estimable 
\cite{Burgarth 2009 2,Burgarth 2011, Maruyama 2012}, and second, it 
is possible to perform a universal gate operation on an infective system 
\cite{Burgarth 2009 3}. 
Moreover, it was proven in \cite{Burgarth PRL 2012,Guta Yamamoto 2012} 
that an infective system is identifiable in the sense discussed in 
Sec.~3.5. 
Therefore, here we are interested in proving the following conjecture:

{\bf Conjecture 4:} 
The controlled SME of an infective system does not have a steady state other 
than the spin coherent state.

By proving the above conjecture and further Conjectures~1 and 2 stated in 
Secs.~3.5 and 4.5, together with the above three facts, we can conclude that 
Conjecture~3 is true, although a more precise meaning of the ``best" structure 
should be clarified.

Lastly, we remark that, in both the identification and stabilization problems, 
the total computation time for running the algorithm largely increases with 
the size of the network; 
so both schemes are inefficient for exponentially large systems. 
In this sense, for instance quantum communication or metrology formulated 
within the indirect control framework, would be suitable subjects to which 
our method should be first applied.

%%%%%%%%%%%%%%%% Acknoledgement %%%%%%%%%%%%%%%%

\ack
This work was supported by JSPS Grant-in-Aid No. 40513289.

%%%%%%%%%%%%%%%%%%%%%%%%%%%%%%%%%%%%%%%%%%%%%%%%%%%%%%%%%%%%%%%%%%
%%%%%%%%%%%%%%%%%%%%%%%%%%%% appendices %%%%%%%%%%%%%%%%%%%%%%%%%%
%%%%%%%%%%%%%%%%%%%%%%%%%%%%%%%%%%%%%%%%%%%%%%%%%%%%%%%%%%%%%%%%%%

\appendix

\section{Derivation of the SME \eref{eq_chase rho} and \eref{eq_chase p}}
\label{Appendix A}

Here we derive the SME \eref{eq_chase rho} and \eref{eq_chase p}, 
using two methods. 
We refer to \cite{Chase 2009,Ralph 2011, Molmer 2013} for more detailed 
description.

The key idea of the first approach is that, by embedding the classical 
probability distribution $\{p_t^{(i)}\}$ into a space of density 
matrices, we apply the quantum filtering theory to the augmented system 
composed of the classical (fictitious) system and the quantum system. 
For this purpose, let $\{ \ket{\psi_i} \}$ be the set of $m$-dimensional 
orthonormal vectors, with the index $i$ corresponding to the $i$-th graph. 
Then, the state of the augmented system is represented by 
\[   \rho^{E}_{t} 
      = \sum_{i=1}^m p^{(i)}_t 
            \ket{\psi_i} \bra{\psi_i} \otimes \rho_t^{(i)}.
\]
In the same manner, the Hamiltonian and the measurement operator 
acting on the whole space are respectively given by 
\[
    H^E = \sum_{i=1}^m \ket{\psi_i} \bra{\psi_i} \otimes H^{(i)}, ~~~ 
    c^E=I_m\otimes c, 
\]
with $c$ given by Eq.~\eref{measurement operator}. 
Then, the SME for the augmented system is given by
\begin{eqnarray}
\label{hybrid sme rho}
& & \hspace*{0em}
    d \rho^{E}_{t} 
      = -i[H^{E}, \rho^{E}_{t}]dt 
          + \gamma \mathcal{D}[c^{E}]\rho^{E}_{t}dt
             + \sqrt{\gamma} \mathcal{H}[c^{E}]\rho^{E}_{t}dW^{E}_{t}, 
\\ & & \hspace*{0em}       
\label{hybrid sme Y}
     dY_{t} 
        = 2\sqrt{\gamma}\hspace{0.05cm}{\rm Tr}(c^E\rho^{E}_{t})dt 
           + dW^{E}_t.       
\end{eqnarray}
In particular, in the basis $\ket{\psi_i}$, Eq.~\eref{hybrid sme rho} gives 
\begin{eqnarray}
\label{hybrid sme p rho_i}
& & \hspace*{-3em}
        dp^{(i)}_{t}\rho^{(i)}_{t} + p^{(i)}_{t} d\rho^{(i)}_{t} 
                   + dp^{(i)}_{t}d\rho^{(i)}_{t}
           = p_t^{(i)}\Big\{ -i[H^{(i)}, \rho^{(i)}_{t}]dt 
                       + \gamma \mathcal{D}[c]\rho^{(i)}_{t}dt 
\nonumber \\ & & \hspace*{4em}
        \mbox{}
             + \sqrt{\gamma} \big[c\rho_t^{(i)} + \rho_t^{(i)}c 
                             - 2\Tr(c^E\rho_t^E)\rho_t^{(i)} \big] dW^{E}_{t} \Big\}. 
\end{eqnarray}
Then, the trace operation on the above equation yields  
\begin{equation}
\label{hybrid sme p}
    dp^{(i)}_{t} 
      = 2\sqrt{\gamma} \big\{ \Tr(c \rho^{(i)}_{t})
          - \Tr(c^E \rho^{E}_{t}) \}p^{(i)}_{t}dW^{E}_{t},
\end{equation}
which is Eq.~\eref{eq_chase p}, where we have expressed 
$\Tr(c^E \rho^{E}_{t})=\Tr(c \tilde\rho_{t})$ with 
$\tilde\rho_t=\sum_i p_t^{(i)} \rho_t^{(i)}$. 
To obtain the equation of $\rho^{(i)}_{t}$, we assume that it follows 
$d\rho^{(i)}_{t} = A_{i} dt + B_{i} dW^{E}_{t}$; 
then substituting this equation and Eq.~\eref{hybrid sme p} for 
Eq.~\eref{hybrid sme p rho_i}, we have $B_{i} = \mathcal{H}[c]\rho^{(i)}_{t}$ 
and 
\[
\hspace{-3em}
     A_{i} = -i[H^{(i)}, \rho^{(i)}_{t}]dt
		+ \gamma \mathcal{D}[c]\rho^{(i)}_{t}dt
			+ 2\sqrt{\gamma} \big\{ \Tr(c^E \rho^{E}_{t})
					- \Tr(c \rho^{(i)}_{t}) \} \mathcal{H}[c]\rho^{(i)}_{t}.
\]
Hence we obtain Eq.~\eref{eq_chase rho}.

Next, we give an alternative way to derive Eqs.~\eref{eq_chase rho} 
and \eref{eq_chase p}. 
The idea is that we explicitly use the classical Bayes rule to obtain the 
update law of $p_t^{(i)}$, unlike the above approach where the Bayes 
rule was implicitly used through the filtering procedure. 
Let us begin with the assumption that, through the estimation process 
up to time $t$, we have estimated the true graph to be $G^{(i)}$; 
then, under this condition, the measurement result is given by 
$dY_t = 2\sqrt{\gamma}\Tr(c\rho_t^{(i)})dt+dW_t$. 
With this information, the conditional state $\rho_t^{(i)}$ is updated 
to $\rho_{t+dt}^{(i)}$ through the usual quantum filtering technique, 
which leads to Eq.~\eref{eq_chase rho}. 
Moreover, it is used for updating 
$p_t^{(i)}={\mathbb P}(\{ G=G^{(i)} \}\hspace{0.1em}|\hspace{0.1em}{\cal Y}_t)$ 
via the Bayes rule ($\mathcal{N}$ is the normalization constant): 
\begin{eqnarray}
& & \hspace*{-5em}
    p_{t+dt}^{(i)} 
      = \frac{
            {\mathbb P}({\cal Y}_{t+dt}\hspace{0.1em}|\hspace{0.1em}\{ G=G^{(i)} \})
                p_{t}^{(i)} }
            { \sum_i 
              {\mathbb P}({\cal Y}_{t+dt}\hspace{0.1em}|\hspace{0.1em}\{ G=G^{(i)} \})
                p_{t}^{(i)}}
      = \frac{1}{\mathcal{N}}
            {\rm exp}\Big[ -\frac{1}{2dt}
                 \Big(dY_t-2\sqrt{\gamma}\Tr(c\rho_t^{(i)})dt\Big)^2
                               \Big]p_{t}^{(i)}
\nonumber \\ & & \hspace*{-3em}
      = \frac{ \Big( 1 + 2\sqrt{\gamma}\Tr(c\rho_t^{(i)})dY_t \Big) p_{t}^{(i)} }
                  { 1 + 2\sqrt{\gamma}\sum_i\Tr(c\rho_t^{(i)}) p_{t}^{(i)} dY_t }
      = \frac{ \Big( 1 + 2\sqrt{\gamma}\Tr(c\rho_t^{(i)})dY_t \Big) p_{t}^{(i)} }
                  { 1 + 2\sqrt{\gamma}\Tr(c\tilde\rho_t) dY_t }
 \nonumber \\ & & \hspace*{-3em}
      = p_{t}^{(i)} 
          + 2\sqrt{\gamma} \Big\{ 
                \Tr (c \rho^{(i)}_{t}) - \Tr (c\tilde{\rho}_{t}) \Big\} 
                      ( dY_{t} - 2\sqrt{\gamma}\Tr( c \tilde{\rho}_{t})dt ) p_{t}^{(i)}, 
\nonumber
\end{eqnarray}
hence we have Eq.~\eref{eq_chase p}.

Here we remark that in reality the output $Y_t$ is generated from the true 
system, thus from Eq.~\eref{true sme Y} the innovation term is given by 
\[
       dW'_t = dW_t + 2\sqrt{\gamma}\hspace{0.05cm}{\rm Tr}(c\rho_t^{(i_0)})dt 
                               - 2\sqrt{\gamma}\hspace{0.05cm}{\rm Tr}(c\rho_t^{(i)})dt,
\]
which is not the standard Wiener increment when $\rho_t^{(i)}\neq \rho_t^{(i_0)}$. 
As a result, particularly when the graph $G^{(i)}$ largely differs from 
the true one $G^{(i_0)}$, the drift term of Eq.~\eref{eq_chase rho} (the term 
proportional to $dt$) can take a big number such that the constraint 
${\rm Tr}(\rho)=1$ or $\rho \geq 0$ is numerically violated; 
consequently in the simulation the time evolution of $\rho_t^{(i)}$ becomes 
unstable and it sometimes diverges. 
Thus, we have introduced a normalization operation in the simulator 
(MATLAB) for numerically preserving those constraints. 
%Now, the scheme shown in \cite{Chase 2009} can be regard as a modification 
%of the hybrid SME so that it evades such computational instability. 
%The idea is that the innovation term $dW'_t$ shown above is replaced by 
%$dW_t + 2\sqrt{\gamma}[{\rm Tr}(c\rho_t^{(i_0)}) - {\rm Tr}(c\tilde\rho_t)]dt$; 
%due to this ensemble averaging of the subtraction term, it turns out that 
%the drift term tends to take a relatively small number, and consequently 
%the numerical simulation becomes stable. 

\section{Steady state of the SME}
\label{Appendix B}

In general, a pure state $\ket{\psi}$ is a steady state of the 
SME \eref{eq_sme1} if and only if $\ket{\psi}$ is a common 
eigenvector of $iH+c^\dagger c/2$ and $c$, which can be directly proved 
using the results \cite{Kraus,Yamamoto}. 
Now, $\ket{0^{\otimes N}}$ is clearly an eigenvector of $c=\sigma_1^z$. 
Also noting the relation 
$(\sigma^x\otimes \sigma^x + \sigma^y\otimes \sigma^y)\ket{00}=0$, 
we readily have $H\ket{0^{\otimes N}}=0$. 
Therefore, $\ket{0^{\otimes N}}$ is a common eigenvector of 
$iH+c^\dagger c/2$ and $c$, hence it is a steady state of the SME. 
Note that the above fact does not mean that $\ket{0^{\otimes N}}$ 
is a unique steady state of the SME.

\section{Proof of Theorem 2}
\label{Appendix C}

The goal is to prove that the controlled SME having permutation 
symmetry property has a steady state other than $\ket{0^{\otimes N}}$. 
This can be achieved by showing that, based on the fact mentioned 
in Appendix~B, there exists a common eigenstate of $c$ and $H$ such 
that $c\ket{\phi}=\ket{\phi}$ and $\ket{\phi}\neq \ket{0^{\otimes N}}$. 
Note that the eigenstate satisfying $c\ket{\phi}=-\ket{\phi}$ cannot 
be a steady state due to the adaptive measurement mechanism.

First, let $P$ be a permutation matrix exchanging the indices $0$ 
and $1$ of two specific spins, which however does not act on the 
first node. 
Then, $J_z=\Sigma_{j=1}^{N}\sigma_j^z$ satisfies 
$[P,J_z]=[H,J_z]=[c,J_z]=0$, since $P, H$, and $c$ preserve the total 
$z$ component of the network. 
Thus, $P, H$, and $c$ can be block-diagonalized into $N+1$ blocks 
corresponding to the eigenspaces of $J_z$; that is, 
$P = \textrm{diag}(P_0,...,P_N)$, $H = \textrm{diag}(H_0,...,H_N)$, 
and $c = \textrm{diag}(c_0,...,c_N)$, where the $j$th component acts 
on the subspace spanned by the states with $j$ 
excitations (i.e., the states composed of $j$ spins with $\ket{1}$ 
and $N-j$ spins with $\ket{0}$). 
In particular, the subspace corresponding to $j=1$ is spanned by 
$\ket{100\ldots 0},~\ket{010\ldots 0}, \ldots,~\ket{000\ldots 1}$. 
Now remove $\ket{100\ldots 0}$ from this space and define 
\[
    S_1 = \textrm{span}\Big\{
           \ket{010\ldots 0},~\ket{001\ldots 0},~\ldots,~\ket{000\ldots 1}
            \Big\}. 
\]
Then, any state $\ket{\phi}\in S_1$ satisfies $c\ket{\phi}=\ket{\phi}$. 
Also note that $\ket{0^{\otimes N}}\notin S_1$.

Let us now define $P' = \textrm{diag}(I_1,P_1,I_{_{N}C_{2}},\ldots,I_1)$; 
then $P'$ must have two eigenvalues $\pm 1$, implying that there exists 
an eigenstate $\ket{\phi}\in S_1$ satisfying $P'\ket{\phi}=-\ket{\phi}$. 
We here use the assumption that $H$ is permutation symmetric, which 
leads to $[H, P]=0$ and further $[H, P']=0$. 
Then, from $[H, P']\ket{\phi}=0$ we have 
\[
    P'(H\ket{\phi})=-(H\ket{\phi}), 
\]
which implies $H\ket{\phi}\in S_1$. 
Noting that this relation holds for any state satisfying 
$P'\ket{\phi}=-\ket{\phi}$, we can conclude that $H$ has an 
eigenstate $\ket{\phi'}\in S_1$. 
Together with the fact that $c\ket{\phi'}=\ket{\phi'}$ and 
$\ket{\phi'}\neq \ket{0^{\otimes N}}$, we obtain the assertion. 
$~\blacksquare$

In the case of the five-spins network shown in Fig.~\ref{fig_6}~(e), 
the steady state $\ket{\phi'}$ other than the target, the existence 
of which is shown in the above proof, can be found as follows. 
First, the operators acting on the space $S_1$, i.e. $P_1$, $H_1$, and 
$c_1$, have the following matrix representation:
\[
\hspace{-2cm}
    P_1
      = \left(\begin{array}{cccc}
            0 & 0 & 0 & 1 \\
            0 & 0 & 1 & 0 \\
            0 & 1 & 0 & 0 \\
            1 & 0 & 0 & 0 \\
       \end{array}\right),~~~
    H_1 = \left(\begin{array}{cccc}
            0  & 2.4 &  0  &  0  \\
           2.4 &  0  & 1.8 &  0  \\
            0  & 1.8 &  0  & 2.4 \\
            0  &  0  & 2.4 &  0  \\
           \end{array} \right),~~~
   c_1 = \left(\begin{array}{cccc}
             1 & 0 & 0 & 0 \\
             0 & 1 & 0 & 0 \\
             0 & 0 & 1 & 0 \\
             0 & 0 & 0 & 1 \\
        \end{array} \right). 
\]
The eigenvector of $P_{1}$ corresponding to the eigenvalue $-1$ 
is of the form $(x, y, -y, -x)^\top$. 
This further becomes an eigenvector of $H_1$, if it takes the form 
$(1, a, -a, -1)^\top/\sqrt{2(1+a^{2})}$ with 
$a = (-3 \pm \sqrt{73})/8$, which is of course an eigenvector of 
$c_1$. 
Then we find that it has the following representation in the whole 
system space: 
\begin{equation*}
     \ket{\phi'} 
        = \frac{1}{\sqrt{2(1+a^{2})}}
            \Big( \ket{00001} + a \ket{00010} 
               - a \ket{00100} - \ket{01000}\Big). 
\end{equation*}
%

%%%%%%%%%%%%%%%%%%%%%%%%%%%%%%%%%%%%%%%%%%%%%%%%%%%%%%%%%%%%%%%%%%%
%%%%%%%%%%%%%%%%%%%%%%%%%%%%%%  Ref.  %%%%%%%%%%%%%%%%%%%%%%%%%%%%%
%%%%%%%%%%%%%%%%%%%%%%%%%%%%%%%%%%%%%%%%%%%%%%%%%%%%%%%%%%%%%%%%%%%


\section*{References}



\begin{thebibliography}{}


\bibitem{Divincenzo 2000}
Divincenzo D P 2000 
%The physical implementation of quantum computation, 
{\it Fortschr. Phys.} \textbf{48} 771

\bibitem{Nielsen Book}
Nielsen M and Chuang I 2000 
{\it Quantum Computation and Quantum Information} 
(Cambridge University Press) 

\bibitem{Deutsch 1989}
Deutsch D 1989 
%Quantum computational networks, 
{\it Proc. Roy. Soc. Lond.} A \textbf{425} 73

\bibitem{DiVincenzo 1995}
DiVincenzo D P 1995 
%Two-bit gates are universal for quantum computation,
{\it Phys. Rev.} A \textbf{50} 1015

\bibitem{Lloyd 1995}
Lloyd S 1995 
%Almost any quantum logic gate is universal, 
{\it Phys. Rev. Lett.} \textbf{75} 346

\bibitem{Barenco 1996}
%A. Barenco, C. H. Bennett, R. Cleve, D. P. Divincenzo, 
%N. Margolus, P. Shor, T. Sleator, J. Smolin, and H. Weinfurter, 
Barenco A {\it et al} 1996 
%Elementary gates for quantum computation, 
{\it Phys. Rev.} A \textbf{52} 3457






\bibitem{Burgarth 2009 1}
Burgarth D, Maruyama K and Nori F 2009
%Coupling strength estimation for spin chains despite 
%restricted access, 
{\it Phys. Rev.} A \textbf{79} 020305(R)

\bibitem{Burgarth 2009 2}
Burgarth D and Maruyama K 2009
%Indirect Hamiltonian identification through a small gateway, 
{\it New J. Phys.} \textbf{11} 103019 

\bibitem{Franco 2009}
Franco C D, Paternostro M and Kim M S 2009 
%Hamiltonian tomography in an access-limited setting without 
%state initialization, 
{\it Phys. Rev. Lett.} \textbf{102} 187203

\bibitem{Burgarth 2011}
Burgarth D, Maruyama K and Nori F 2011 
%Indirect quantum tomography of quadratic Hamiltonians, 
{\it New J. Phys.} \textbf{13} 013019

\bibitem{Lapasar 2012}
Lapasar E H, Maruyama K, Burgarth D, Takui T, Kondo Y 
and Nakahara M 2012
%Estimation of coupling constants of a three-spin chain: a case 
%study of Hamiltonian tomography with nuclear magnetic resonance, 
{\it New J. Phys.} \textbf{14} 013043 

\bibitem{Maruyama 2012}
Maruyama K, Burgarth D, Ishizaki A, Takui T and Whaley K B  2012  
%Application of indirect Hamiltonian tomography to complex systems 
%with short coherence times,
{\it Quantum Inf. Comput.} \textbf{12} 763





\bibitem{Nakazato 2003}
Nakazato H, Takazawa T and Yuasa K 2003 
%Purification through zeno-like measurements, 
{\it Phys. Rev. Lett.} \textbf{90} 060401

\bibitem{Nakazato 2004}
Nakazato H, Unoki M and Yuasa K 2004
%Preparation and entanglement purification of qubits through 
%zeno-like measurements, 
{\it Phys. Rev.} A \textbf{70} 012303

%\bibitem{Burgarth 2008}
%Burgarth D and Giovannetti V 
%2008 
%Proc. Quantum Information and Many body Quantum Systems (Pisa,
%Edizioni della Normale) ed M Ericsson and S Montangero p 17 
%(arXiv:0710.0302)

%\bibitem{Wang 2010}
%X. Wang, A. Bayat, S. G. Schirmer, and S. Bose, 
%Robust entanglement in antiferromagnetic Heisenberg chains 
%by single-spin optimal control, 
%Phys. Rev. A \textbf{81}, 032312 (2010).





\bibitem{Schirmer 2008}
Schirmer S G, Kandasamy G and Devitt S J 2008
%Control paradigms for quantum engineering, 
Proceedings of IEEE ISCCSP, Malta 
%p. 966.

\bibitem{Schirmer 2008 2}
Schirmer S, Pullen I and Pemberton-Ross P 2008 
%Global controllability with a single local actuators, 
{\it Phys. Rev.} A \textbf{78} 062339

\bibitem{Burgarth 2009 3}
Burgarth D, Bose S, Bruder C and Giovannetti V 2009 
%Local controllability of quantum networks, 
{\it Phys. Rev.} A \textbf{79} 060305(R)

\bibitem{Burgarth 2010}
Burgarth D, Maruyama K, Murphy M, Montangero S, 
Calarco T, Nori F and Plenio M B 2010
%Scalable quantum computation via local control of only two qubits, 
{\it Phys. Rev.} A \textbf{81} 040303(R)

\bibitem{Wang 2012}
Wang X, Pemberton-Ross P and Schirmer S G 2012
%Symmetry and subspace controllability for spin networks 
%with a single-node control, 
{\it IEEE Trans. Autom. Control} \textbf{57} 1945





\bibitem{Belavkin}
Belavkin V P 1992
%Quantum stochastic calculus and quantum nonlinear filtering, 
{\it J. Multivariate Anal.} \textbf{42} 171/201

\bibitem{Bouten}
Bouten L, van Handel R and James M R 2007
%An introduction to quantum filtering, 
{\it SIAM J. Control Optim.} \textbf{46} 2199/2241 

\bibitem{Wiseman Book}
Wiseman H M and Milburn G J 2009 
{\it Quantum Measurement and Control}
(Cambridge Univ. Press) 




\bibitem{Mabuchi 1996}
Mabuchi H 1996
%Dynamical identification of open quantum systems, 
{\it Quantum Semiclassic. Opt.} \textbf{8} 1103

\bibitem{Gambetta 2001}
Gambetta J and Wiseman H M 2001
%State and dynamical parameter estimation for open quantum systems, 
{\it Phys. Rev.} A \textbf{64} 042105

\bibitem{Chase 2009}
Chase B A and Geremia JM 2009
%Single-shot parameter estimation via continuous quantum measurement, 
{\it Phys. Rev.} A \textbf{79} 022314

\bibitem{Ralph 2011}
Ralph J F, Jacobs K and Hill C D 2011 
%Frequency tracking and parameter estimation for robust quantum state estimation, 
{\it Phys. Rev.} A \textbf{84} 052119

\bibitem{Molmer 2013}
Gammelmark S and Molmer K 2013
%Bayesian parameter inference from continuously monitored quantum systems, 
{\it Phys. Rev.} A \textbf{87} 032115




\bibitem{Stockton 2004}
Stockton J K, van Handel R and Mabuchi H 2004
%Deterministic Dicke-state preparation with continuous measurement 
%and control, 
{\it Phys. Rev.} A \textbf{70} 022106

\bibitem{Handel 2005}
van Handel R, Stockton J K and Mabuchi H 2005
%Feedback control of quantum state reduction, 
{\it IEEE Trans. Autom. Control} \textbf{50} 768/780 

\bibitem{Yamamoto 2007}
Yamamoto N, Tsumura K and Hara S 2007
%Feedback control of quantum entanglement in a two-spin system, 
{\it Automatica} \textbf{43} 981/992

\bibitem{Mirrahimi}
Mirrahimi M and van Handel R 2007
%Stabilizing feedback controls for quantum systems, 
{\it SIAM J. Control Optim.} \textbf{46} 445/467 








%\bibitem{Gladwell}
%G. M. L. Gladwell, 
%{\it Inverse Problems in Vibration}, 
%(Dordrecht: Kluwer) 2004

%18
\bibitem{Rice 2005}
Rice J J, Tu Y and Stolovitzky G 2005 
%Reconstructing biological networks using conditional 
%correlation analysis, 
{\it Bioinformatics} \textbf{21} 765

%19
%\bibitem{Snijders 2010}
%T. A. B. Snijders, J. Koskinen, and M. Schweinberger, 
%Maximum likelihood estimation for social network dynamics, 
%Annals of Applied Statistics \textbf{4}, 567 (2010).

%20
\bibitem{Siciliano 2012}
Siciliano M D, Yenigunc D and Ertanb G 2012 
%Estimating network structure via random sampling: 
%Cognitive social structures and the adaptive threshold method, 
{\it Social Networks} \textbf{34} 585 







\bibitem{Jacobs 2010}
Jacobs K 2010
%Feedback control using only quantum back-action, 
{\it New J. Phys.} \textbf{12} 043005 


\bibitem{Tanaka 2012}
Tanaka S and Yamamoto N 2012
%Robust adaptive measurement scheme for qubit-state preparation, 
{\it Phys. Rev.} A \textbf{86} 062331 








%\bibitem{Christandl 2004}
%M. Christandl, N. Datta, A. Ekert, and A. J. Landahl, 
%Perfect state transfer in quantum spin networks, 
%Phys. Rev. Lett. \textbf{92}, 187902 (2004).

\bibitem{Lieb1961}
Lieb E, Schultz T and Mattis D 1961
%Two soluble models of an antiferromagnetic chain, 
{\it Annals of Physics} \textbf{16} 407/466 






\bibitem{Bloembergen}
Bloembergen N 1949
%On the interaction of nuclear spins in a crystalline lattice, 
{\it Physica} \textbf{15} 386/426

\bibitem{Negoro}
Negoro M, Tateishi K, Kagawa A and Kitagawa M 2011
%Scalable spin amplification with a gain over a hundred, 
{\it Phys. Rev. Lett.} \textbf{107} 050503




\bibitem{LjungBook}
Ljung L 1987
{\it System Identification: Theory for the User}
(Prentice Hall)

\bibitem{Burgarth PRL 2012}
Burgarth D and Yuasa K 2012
%Quantum system identification, 
{\it Phys. Rev. Lett.} \textbf{108} 080502

\bibitem{Guta Yamamoto 2012}
Guta M and Yamamoto N 2013
%Systems identification for passive linear quantum systems:
%the transfer function approach, 
arXiv:1303.3771; 
Proceedings of 52nd IEEE CDC 




\bibitem{Burgarth 2007}
Burgarth D and Giovannetti V 2007 
%Full control by locally induced relaxation, 
{\it Phys. Rev. Lett.} \textbf{99} 100501 




\bibitem{Hasminskii}
Has'minskii R Z 1980
{\it Stochastic stability of differential equations} 
(Alphen a/d Rijn: Sijthoff $\&$ Noordhoff) 

\bibitem{Kraus}
Kraus B, B\"uchler H P, Diehl S, Kantian A, Micheli A 
and Zoller P 2008
%Preparation of entangled state by quantum Markov processes, 
{\it Phys. Rev.} A \textbf{78} 042307

\bibitem{Yamamoto}
Yamamoto N 2005
%Parametrization of feedback Hamiltonian realizing a pure steady 
%state, 
{\it Phys. Rev.} A \textbf{72} 024104

\end{thebibliography}
\end{document}